\documentclass[fleqn,twoside]{article}
\usepackage{espcrc2}

\usepackage{epsfig}
\usepackage[figuresright]{rotating}

\newcommand{\bce}{\begin{center}} 
\newcommand{\ece}{\end{center}}
\newcommand{\beq}{\begin{equation}}
\newcommand{\eeq}{\end{equation}}
\newcommand{\bea}{\vspace{0.25cm}\begin{eqnarray}}
\newcommand{\eea}{\end{eqnarray}}

\newcommand{\br}{{\bf r}}
\newcommand{\bR}{{\bf R}}

\newcommand{\ba}{\begin{array}}
\newcommand{\ea}{\end{array}}

\newcommand{\bfrho}{\mbox{\boldmath $\rho$}}

\newcommand{\bkappa}{\mbox{\boldmath ${\kappa}$}}

\def\lsim{\mathrel{\rlap{\lower4pt\hbox{\hskip1pt$\sim$}}
    \raise1pt\hbox{$<$}}}         %less than or approx. symbol
\def\gsim{\mathrel{\rlap{\lower4pt\hbox{\hskip1pt$\sim$}}
    \raise1pt\hbox{$>$}}}         %greater than or approx. symbol

\def\Pom{{\bf I\!P}}

\def\beq{\begin{equation}}
\def\endeq{\end{equation}}
\def\arr{\begin{eqnarray}}
\def\endarr{\end{eqnarray}}

\newcommand{\AmS}{{\protect\the\textfont2
  A\kern-.1667em\lower.5ex\hbox{M}\kern-.125emS}}

\title{Triple-pomeron dynamics: valence vs. glue}

\author{V.R.~Zoller\address {ITEP, Moscow 117218, Russia}}

\begin{document}

\begin{abstract}
Three pomerons are known to couple  via the gluon loop and 
interactions of this kind which are responsible for high-mass diffraction
are  described in terms of the gluonic structure
function of the QCD pomeron. We show that the triple-pomeron  coupling 
via the light quark loop associated with the ``valence'' in the pomeron
is of the same strength as the purely gluonic coupling.     
The large $Q^2$ behavior of this new contribution is described 
by the DLLA evolution from the non-perturbative $f\bar f$ valence state of 
the pomeron. Numerical estimates of the high-mass diffraction structure 
functions based on the consistent account of both couplings 
 are in good agreement with experimental data.
\vspace{1pc}
\end{abstract}

\maketitle

In this communication  we  discuss
the new triple-pomeron
{\footnote{for definitions and early analysis see \cite{KTM}}} 
contribution to the  high-mass diffractive 
deep inelastic scattering (DIS)   which 
comes from excitation of the $(q\bar{q})(f\bar{f})$ Fock states of the photon,
\beq
\gamma^*+p \to (q\bar{q})(f\bar{f})+p
\label{eq:QQFF}
\eeq
where $q$ and $f$ are the light quarks \cite{NSZZ}. 
Compared to the purely gluonic triple pomeron interactions which scale is
determined by the correlation radius of non-perturbative gluons, $R_c\sim 0.25$
fermi \cite{GNZ95DifDIS}, the $f\bar{f}$-pair in the quark loop 
connectig pomerons has a large non-perturbative size,
$r_f \simeq 1/m_f\sim 1$ fermi,
 and  enters the evolution of  sea of the pomeron
in precisely the same way as the valence quark density 
enters the evolution of the sea of nucleons. 
We  demonstrate that 
 such a pQCD
evolution holds at least to the DLLA accuracy. We show that
although the $(q\bar{q})(f\bar{f})$ contribution is of higher order
in pQCD coupling $\alpha_S$  it  
is enhanced by a potentially large numerical factor, 
$\propto \left[\sigma(r_f)/\sigma(R_c)\right]^2$, where 
$\sigma(r)\sim r^2$ is the color dipole cross section, and numerically
it is comparable to the leading order gluonic triple-pomeron term.

The process (\ref{eq:QQFF}) represents the first step of 
$(\beta,Q^2)$-evolution of the diffractive structure function 
from  the $f\bar f$ ``valence'' quark component of 
the pomeron. Its space-time picture in the lab. reference frame is as follows.
  The high-energy  photon converts into the  colorless $q\bar q$
pair at large distances upstream the target.
The $q\bar q$ dipole  with  transverse
separation ${\bf r}$  acts like  a source of gluons. 
Emitted gluon has 
a transverse coordinate $\bfrho$ and the  momentum fraction $z_g$
which are distributed in accord with  the wave 
function of the $q\bar q g$ state \cite{NZ3P}. This gluon has a virtuality 
$\kappa^2\simeq -{\bkappa}^2$ and in its turn converts  into
the  pair of quarks $f\bar f$ of size ${\bf R}$. 
To  isolate the leading  terms  of  the 4-parton interaction 
cross section one has  to look into  the impact parameter structure 
of the diffractive 
$(q\bar q)(f\bar f)$ state. The analysis of ref.\cite{NZ92} shows 
 that the effective 
 size of the quark-anti-quark component of the photon is generally 
 process-dependent. In particular, the proton structure function 
$F_{2p}(x,Q^2)$ in the scaling limit is dominated by the small-size 
$(q\bar q)$-fluctuations,
$
r^2\sim {1/Q^2}.
$
The case of the diffraction dissociation (DD) of the photon is quite different.
The $q\bar q$ fluctuations of the size $1/Q^2$
are altogether negligible and  the DD structure function 
is dominated by the $q\bar q$ states
of the largest possible
size $R\sim 1/m_q\sim 1$ fermi. 
Thus, there are two different scattering regimes and two very different 
scales. Both of them are important for 
the process (\ref{eq:QQFF}) where the two quark pairs,  $q\bar q$
and  $f\bar f$, are in the regimes of inclusive DIS and DD,
respectively. 
As we are interested in the large mass excitations, 
$M^{2} \propto Q^{2}/z_{g}$, the Sudakov variable for the gluon is 
$z_g\ll 1$. If so, the small size  $q\bar q$-dipole and 
radiated gluon $g$ are separated from the origin 
(which is fixed by the projectile photon) by the distances $z_g\rho$ and 
$(1-z_g)\rho$, respectively. The $f\bar f$-pair produced by the gluon with
virtuality  $\kappa^2\gsim R_c^{2}$ is in the DD regime and, as such,
is very asymmetric. Its Sudakov variable is
$
\alpha\lsim {m_f^2/\kappa^2}.
$
Two conclusions can be drawn immediately. First, the size of $f\bar f$-pair
is large,
$ 
R\sim {1/m_f},
$ 
and, second, the $f$-(anti)quark  is separated from the origin by the 
large distance
$\sim R(1-{m_f^2/\kappa^2})$,
while the other (anti)quark goes along the parent gluon
with small separation $\sim R {m_f^2/\kappa^2}$.
Hence, the DLLA ordering of  sizes \cite{DGLAP}
which vary along the partonic  ladder from very large 
$R\sim 1/m_f$ in the bottom  ${f\bar f}$-cell 
of the ladder down to very small $r\sim 1/\sqrt{Q^2}$ in the top, $q\bar q$- 
cell, 
\beq
 r\ll \rho \ll R.
\label{eq:DLLA}
\eeq
As a result the 4-parton system 
$(q\bar q)(f\bar f)$ acts like a triplet-anti-triplet dipole of size 
$R\sim 1/m_f\sim 1$ fermi 
and as such  has large interaction cross section 
$  
\sigma_{(q\bar q)(f\bar f)}\simeq \sigma(R).
$
 The distribution of $f\bar{f}$ color 
dipoles in the gluon of transverse momentum $\bkappa$ is 
identical to that in the photon subject to the substitutions
$N_c\alpha_{em} e_f^2\to T_F\alpha_S(\bkappa^2)$ and $Q^2 \to \bkappa^2$, 
so that the diffractive cross section of interest equals
\bea
(Q^{2}+M^{2})\left.{d\sigma^{D}((q\bar q)(f\bar f)) 
\over dt dM^{2}}\right|_{t=0}=
{1\over 16\pi}
\int d^2\bkappa\nonumber\\
\times {dg_{q\bar{q}}(Q^2,\bkappa)\over d^2\bkappa}
 {T_F\alpha_S(\bkappa^2)\over N_c\alpha_{em} e_f^2} 
\langle f\bar{f}|\sigma^2(x_{\Pom},\bR)|f\bar{f}\rangle\, ,
\label{eq:qqff}
\eea
where the flux of gluons in the parent $q\bar{q}$ state is 
\bea
{dg_{q\bar{q}}(Q^2,\bkappa)\over d^2\bkappa} =
\int_0^1 dz_q
\int d^2{\bf r}\left|\Psi_{\gamma^*}(Q^2,z_q,{\bf r})\right|^2\nonumber\\
 \times{2e_q^2C_F\alpha_S(r^2) \over \pi^2}\cdot 
{\left[1-\exp(i\bkappa{\bf r})\right]\over 
(\bkappa^2+\mu_G^2)^2}\bkappa^2 .         
\label{eq:NG}
\eea
Finally, notice that 
\bea
{\bkappa^2\over  4\pi^2 \alpha_{em}} 
{1\over 16\pi e_f^2}\langle f\bar{f}|\sigma^2(x_{\Pom},\bR)|f\bar f\rangle\nonumber\\
={\bkappa^2\over  4\pi^2 \alpha_{em}}
\left.{d\sigma(\gamma^*(\bkappa^2) \to f\bar{f}) \over dt}\right|_{t=0} 
\nonumber\\
= {1\over e_f^2}
\int_{0}^{1} {d\beta \over \beta} 
f^D_{f\bar{f}}(t=0,x_{\Pom},\beta,\bkappa^2)= 
N_{f\bar{f}}^{\Pom}(x_{\Pom},\bkappa^2)
\label{eq:Nff}
\eea
where $N_{f\bar{f}}^{\Pom}(x_{\Pom},\bkappa^2)$ can be reinterpreted as a
number of charged valence partons, i.e., twice the number of $f\bar{f}$ 
dipoles,
in the pomeron. Upon the substitution of (\ref{eq:Nff}) and (\ref{eq:NG})
into (\ref{eq:qqff}) one readily recovers the dipole representation 
\bea
(Q^{2}+M^{2})\left.{d\sigma_{q\bar q g}^{D} 
\over dt dM^{2}}\right|_{t=0} = 
\langle q\bar{q}|\sigma^{\Pom}(x_{\Pom},\br)|q\bar{q}\rangle\,,
\label{eq:DSDT2}
\eea
in which $\sigma^{\Pom}(x_{\Pom},\br)$ is evaluated 
for the
unintegrated gluon structure function evolved from the $f\bar{f}$ state of the
pomeron
$
{\cal F}_{f\bar{f}}^{\Pom}(\beta,\bkappa^2)= (C_F \alpha_S/\pi)  
N_{f\bar{f}}^{\Pom}(x_{\Pom},\bkappa^2).
$
Furthermore,  
$N_{f\bar{f}}^{\Pom}(x_{\Pom},\bkappa^2)$ vanishes at $\bkappa^2 =0$ and,
according to \cite{NZ92}, 
flattens at $\bkappa^2 \gg m_f^2$. 
The DLLA analysis of $q\bar{q}g_1...g_n$ excitation developed in \cite{NZ3P}
can readily be extended to the higher, $(q\bar{q})g_1..g_n(f\bar{f})$, states.
\begin{figure}[h]
\vspace{-0.9cm}
\hspace{-0.5cm}
\psfig{figure=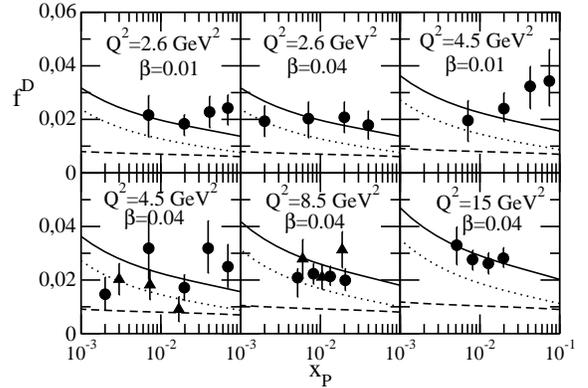,angle=-90,scale=0.33}
\vspace{-1.0cm}
\hspace{-0.5cm}
\caption{The comparison with the experimental data 
on small-$\beta$, small-$x_{\Pom}$ diffractive structure function
(\protect\cite{H1}, full circles;
\protect\cite{ZEUS}, full triangle) 
of the theoretical evaluation of 
$f^{D(3)}=f_{q\bar qg}^{D(3)}+f_{(q\bar q)(f\bar f)}^{D(3)}$ 
shown by the solid line. 
The dotted line corresponds to $f^{D(3)}_{q\bar qg}$ and the dashed line
represents $f_{(q\bar q)(f\bar f)}^{D(3)}$.}
\label{fig:F3D}
\end{figure}  
The numerical results for high-mass, small-$\beta$,  
diffraction depend on the input dipole cross section $\sigma(x,\br)$.
Here we evaluate the lowest order $q\bar{q}g$ \cite{GNZ95DifDIS} and 
$(q\bar{q})(f\bar{f})$ \cite{NSZZ}
contributions to diffractive DIS
\bea
f^{D(4)}(t=0,x_{\Pom},\beta,Q^2)= {Q^2\over 4\pi^2\alpha_{em}}\nonumber\\
\times(Q^{2}+M^{2})\left.{d\sigma^{D} 
\over dt dM^{2}}\right|_{t=0}
\label{eq:1.9}
\eea 
 in a specific  color 
dipole BFKL model \cite{CDBFKL} which gives a good description
of the proton structure function data. The applicability domain of
the small-$\beta$, small-$x_{\Pom}$ formalism is $\beta,x_{\Pom} < x_0 \ll 1$,
the experience with inclusive DIS suggests $x_{0}\sim 0.03$, although
the theoretical curves in fig.~1 are stretched up to $x_{\Pom}=0.1$. 
This small-$\beta$, small-$x_{\Pom}$  
domain is almost at the boundary of the HERA experiments, the 
corresponding experimental data on the $t$-integrated diffractive structure 
function
$f^{D(3)}(x_{\Pom},\beta,Q^2)$ from H1 (\cite{H1}, circles) and ZEUS 
(\cite{ZEUS}, triangles)
are shown in fig.~1. We evaluate this structure function as
$f^{D(3)}(x_{\Pom},\beta,Q^2) = \int dt f^{D(4)}(t,x_{\Pom},\beta,Q^2) \approx
{1\over B_{3\Pom}} f^{D(4)}(t=0,x_{\Pom},\beta,Q^2)$ with the central value of
the diffraction slope $B_{\Pom} =
B_D=7.2\pm 1.1^{+0.7}_{-0.9}$ GeV$^{-2}$ as reported by ZEUS 
\cite{BD}.  The apparent growth of the experimentally observed 
$f^{D(3)}(x_{\Pom},\beta,Q^2)$  towards large $x_{\Pom}\sim 0.1$ is usually
attributed to the non-vacuum admixture to the pomeron exchange.
Two features of the theoretical results for small-$\beta$ 
diffraction are noteworthy. First, the contributions from $q\bar{q}g$ and 
higher-order $(q\bar{q})(f\bar{f})$ states are of comparable magnitude
because $R_c \ll r_f$ and the latter is enhanced 
$\propto [\sigma(x_{\Pom},r_f)/\sigma(x_{\Pom},R_c)]^2$.
Second, because of the same inequality of the important dipole sizes,
$R_c \ll r_f$,
the $x_{\Pom}$-dependence of the $q\bar{q}g$ excitation is steeper 
than that of the  $(q\bar{q})(f\bar{f})$ excitation, 
the numerically significant contribution
from the $(q\bar{q})(f\bar{f})$ excitation makes the overall
$x_{\Pom}$-dependence of $f^{D(3)}(x_{\Pom},\beta,Q^2)$ weaker than evaluated
in for the pure  $q\bar{q}g$ excitation. The solid curve
in fig.~1 is the combined contribution from the two mechanisms. It is
in reasonable agreement with the HERA data.

It is a pleasure to thank Roberto Fiore for his kind invitation and 
warm hospitality at the Diffraction 2004.
This work has been partly supported by the INTAS grant 00-00366 and
the DFG grant 436RUS17/101/04.

\end{document}